\documentclass[]{aa}
\usepackage{graphics,epsfig,txfonts}
\usepackage{natbib}
\bibpunct{(}{)}{;}{a}{}{,}

\usepackage{graphicx}

\usepackage{amssymb}
\usepackage{xcolor}
\usepackage{lipsum}
\usepackage{textcomp}
\usepackage{enumerate}

\newcommand{\oergs}[1]{$10^{#1}$ erg s$^{-1}$}
\newcommand{\hcm}[1]{$\times 10^{#1}$ cm$^{-2}$}

\newcommand{\expo}[1]{$\times 10^{#1}$}
\newcommand{\oexpo}[1]{$10^{#1}$}
\newcommand{\kms}{km s$^{-1}$}
\newcommand{\nh}{N$_{\rm H}$}

\newcommand{\ct}{cts s$^{-1}$}

\newcommand{\ltsima}{$\buildrel < \over \sim$}
\newcommand{\lsim}{\lower.5ex\hbox{\ltsima}}
\newcommand{\gtsima}{$\buildrel > \over \sim$}
\newcommand{\gsim}{\lower.5ex\hbox{\gtsima}}

\newcommand{\swiftp}{\hbox{Swift\,J053041.9-665426}}
\newcommand{\xmm}{{\it XMM-Newton}}

\begin{document}
 
\title{Swift\,J053041.9-665426, a new Be/X-ray binary pulsar in the Large Magellanic Cloud
\thanks{Based on observations with 
               \xmm, an ESA Science Mission with instruments and contributions 
               directly funded by ESA Member states and the USA (NASA)}
}

\author{G.~Vasilopoulos\inst{1} \and  P.~Maggi\inst{1} \and F.~Haberl\inst{1} \and R.~Sturm\inst{1} \and W.~Pietsch\inst{1} \and E.~S.~Bartlett\inst{2} \and M.~J.~Coe\inst{2}  }

\titlerunning{\swiftp, a new Be/X-ray binary pulsar in the LMC}
\authorrunning{Vasilopoulos et al.}
 
\institute{Max-Planck-Institut f\"ur extraterrestrische Physik,
           Giessenbachstra{\ss}e, 85748 Garching, Germany\\
	   \email{gevas@mpe.mpg.de}
	   \and
           School of Physics and Astronomy, University of Southampton, Highfield, Southampton SO17 1BJ, UK
	   }
 
\date{Received ?? ??? 2013 / Accepted ?? ??? 2013}
 
\abstract{}
{We observed the newly discovered X-ray source \swiftp~ in the X-ray and optical regime to confirm its proposed nature as a high mass X-ray binary.}
{We obtained \xmm~  and Swift X-ray data, along with optical observations with the ESO Faint Object Spectrograph, to investigate the spectral and temporal characteristics of \swiftp.}
{The \xmm~ data show coherent X-ray pulsations with a period of $28.77521(10)$ s ($1\sigma$). The X-ray spectrum can be modelled by an absorbed power law with photon index within the range 0.76 to 0.87. The addition of a black body component increases the quality of the fit but also leads to strong dependences of the photon index, black-body temperature and absorption column density. We identified the only optical counterpart within the error circle of \xmm~ at an angular distance of $\sim$0.8 \arcsec, which is \object{2MASS\,J05304215-6654303}. We performed optical spectroscopy from which we classify the companion as a B0-1.5Ve star.}
{The X-ray pulsations and long-term variability, as well as the properties of the optical counterpart, confirm that \object{Swift\,J053041.9-665426} is a new Be/X-ray binary pulsar in the Large Magellanic Cloud.}


\maketitle

\section{Introduction}
\label{sec-intro}

\begin{table*}
 \caption{X-ray observations of \swiftp.}
 \begin{center}
\scalebox{0.8}{   \begin{tabular}{lccccccrrrr}
     \hline\hline\noalign{\smallskip}
     \multicolumn{1}{c}{ObsID} &
     \multicolumn{1}{c}{Date} &
     \multicolumn{1}{c}{Start time} &
     \multicolumn{1}{c}{Instrument} &
     \multicolumn{1}{c}{Mode\tablefootmark{a}} &
     \multicolumn{1}{c}{Offax\tablefootmark{b}} &
     \multicolumn{1}{l}{Net Exp} &  
     \multicolumn{1}{r}{Net Count rates\tablefootmark{c}} &
     \multicolumn{1}{r}{R$_{\rm sc}$\tablefootmark{d}} &
     \multicolumn{1}{c}{$F_{\rm x}$\tablefootmark{e}} &
     \multicolumn{1}{c}{$L_{\rm x}$\tablefootmark{f}}   \\

     \multicolumn{1}{l}{} &
     \multicolumn{1}{l}{} &
     \multicolumn{1}{c}{(UT)} &
     \multicolumn{1}{l}{} &
     \multicolumn{1}{l}{} &
     \multicolumn{1}{c}{[\arcmin]} &
     \multicolumn{1}{c}{[ks]} &   
     \multicolumn{1}{c}{[cts s$^{-1}$]} &
     \multicolumn{1}{r}{[\arcsec]} &
     \multicolumn{1}{c}{[erg s$^{-1}$ cm$^{-2}$]}  &
     \multicolumn{1}{c}{[erg s$^{-1}$]}      \\
     \noalign{\smallskip}\hline\noalign{\smallskip}

            XMM         &  &  &  &  &  &  &  &  &  & \\
            0700381101  &  2012-09-03  &  21:34 -- 07:34   &  EPIC-pn    &  ff--m  &  1.12  &  22.3  & 0.211$\pm$0.003   & 20  &  1.79$\times10^{-12}$  & $5.53\times10^{35}$ \\
                        &              &  21:11 -- 07:34   &  EPIC-MOS1  &  ff--m  &  0.22  &  26.0    & 0.0765$\pm$0.0019   & 32  & & \\
                        &              &  21:11 -- 07:34   &  EPIC-MOS2  &  ff--m  &  1.25  &  26.0    & 0.0758$\pm$0.0018 & 26  &  & \\


         
     \noalign{\smallskip}\hline\noalign{\smallskip}
        Swift         &  &  &  &  &  &  &  &  &  & \\
        00045769001   &  	2011-11-06  &   03:54:01  &  XRT     &  pc     &	 10.4	& 0.4  & 0.35 $\pm$0.03    & 35 & 3.49$\times10^{-11}$ & 1.04$\times10^{37}$ \\     
        00032172001   &  	2011-11-08  &   16:41:01  &  XRT     &  pc     &	 1.6 	& 1.0  & 0.498$\pm$0.023   & 35 & 4.16$\times10^{-11}$ & 1.24$\times10^{37}$ \\
        00032172002   &  	2011-11-10  &   13:38:00  &  XRT     &  pc     &	 1.9 	& 1.2  & 0.427$\pm$0.020   & 35 & 3.07$\times10^{-11}$ & 0.92$\times10^{37}$ \\
        00032172003   &  	2011-11-12  &   13:51:01  &  XRT     &  pc     &	 2.0 	& 1.1  & 0.417$\pm$0.021   & 35 & 2.88$\times10^{-11}$ & 0.86$\times10^{37}$ \\
        00045769002   &  	2012-08-28  &   01:08:58  &  XRT     &  pc     &         10.9   & 1.9  & 0.098$\pm$0.009   & 35 & 9.47$\times10^{-12}$ & 2.83$\times10^{36}$ \\
        00045769004   &  	2012-10-18  &   07:02:34  &  XRT     &  pc     &	 11.9	& 0.5  & 0.033$\pm$0.010   & 35 & 3.02$\times10^{-12}$ & 9.04$\times10^{35}$ \\

     \noalign{\smallskip}\hline\noalign{\smallskip}

        ROSAT         &  &  &  &  &  &  &  &  &  & \\
        RP900553N00  &  1999-11-09  & 15:28 -- 22:26   &  PSPC    & -   &  -   &  1.1  &  <0.006  & - &  <1.58$\times10^{-12}$ & <4.70$\times10^{35}$ \\

      \noalign{\smallskip}\hline\noalign{\smallskip}
    \end{tabular}   }
 \end{center}
  \tablefoot{
  \tablefoottext{a}{Observation setup: full-frame mode (ff) and photon-counting mode (pc). For \xmm, the medium filter (m) was used.}
  \tablefoottext{b}{Off-axis angle under which the source was observed.}
  \tablefoottext{c}{Net counts as used for spectral analysis in the (0.2 -- 10.0) keV band for \xmm\ and in the (0.3 -- 6.0) keV band for {\em Swift}.}
  \tablefoottext{d}{Radius of the circular source extraction region.}
  \tablefoottext{e}{X-ray flux in the (0.3 -- 10.0) keV band, derived from the best fit spectral model.}
  \tablefoottext{f}{Source intrinsic X-ray luminosity in the (0.3 -- 10.0) keV band (corrected for absorption)
for a distance to the LMC of 50 kpc \citep{2005MNRAS.357..304H}.}
  }
 \label{tab:xray-obs}
\end{table*}

Be/X-ray binaries (BeXRBs) are a subclass of high mass X-ray binaries (HMXRBs). The binary consists of a compact object, typically a neutron star, that accretes mass from a high-mass main sequence, subgiant or giant OBe star. Currently, there are no confirmed black hole BeXRB systems \citep{2009ApJ...707..870B}. The optical companions in BeXRBs are rapidly rotating emission-line stars, typically having spectral classes between O5 and B9 \citep{2005MNRAS.356..502C}. The rapidly rotating high mass star loses material resulting in the formation of an equatorial disc, that is also referred to as decretion disc. The exact mechanism by which material is fed by the star to the disc is not yet well understood \citep{2003PASP..115.1153P}.
This disc is responsible for the optical emission lines, mainly the Balmer and Paschen series of Hydrogen and sometimes He and Fe lines. Another signature of these systems is the infrared (IR) excess also originating from the equatorial disc \citep[for a full review see][]{2011Ap&SS.332....1R}.

In a BeXRB system the neutron star is usually orbiting its companion in a significantly eccentric orbit that may be inclined to the decretion disc \citep{2011MNRAS.416.1556T}. This results in higher accretion rates when the stellar separation is smaller and/or the neutron star passes through the decretion disc, which can lead to the X-ray emission being transient in nature. 
During these phases of enhanced accretion the X-ray luminosity can reach values up to \oergs{36-37}. These so called Type I outbursts can be identified from their periodic behaviour related to the orbital period of the binary system. Sometimes, giant outbursts (Type II) may occur at luminosities greater than \oergs{37} and last for several orbital periods. These are believed to be caused by large structural changes in the decretion disc which can be completely lost and replenished \citep{2013PASJ...65...41O}.  

The strong magnetic field of the neutron star funnels mass accretion onto the magnetic poles and leads to the formation of an accretion column. Shocks heat the gas which then emits X-rays (for a detailed 
review of the emission mechanisms see \citealt{2004ApJ...614..881H,2002apa..book.....F}). Due to the localised mass accretion onto the magnetic poles of the rotating neutron star, it is possible to detect pulses in the X-ray light curve. The pulse profiles can be single peaked, double peaked or even more complex depending on the accretion geometry and viewing angle \citep{2010A&A...520A..76A}.    

In the last decade, a thermal soft excess has been detected in several BeXRB pulsars. It has been suggested that the presence of a soft spectral component could be a very common feature intrinsic to X-ray pulsars. \citet{2004ApJ...614..881H} concluded that the origin of this excess could be related to the total luminosity of the source. For sources with higher luminosity ($L_X>$\oergs{38}) the excess could be explained only by reprocessing of hard X–rays from the neutron star by optically thick accreting material. In less luminous sources ($L_X<$\oergs{36}), the soft excess can be due to emission by photoionised or collisionally heated gas or thermal emission from the surface of the neutron star. Either or both of these types of emission can be present for sources with intermediate luminosity.  

The study of BeXRBs in nearby galaxies can provide us with more complete and homogeneous samples of BeXRB populations with the advantage of low foreground absorption and well known distances compared to sources in our Galaxy. The Magellanic Clouds are the closest star-forming galaxies to our own, making them ideal for the study of their X-ray binary populations. 
Approximately 60 confirmed pulsars have been found in the Small Magellanic Cloud (SMC) so far \citep{2004A&A...414..667H,2005MNRAS.356..502C,2010ASPC..422..224C}, and 45 candidates have been identified during the recent \xmm~ survey of the SMC \citep[][]{2013arXiv1307.7594S,2012A&A...545A.128H}. On the other hand the Large Magellanic Cloud (LMC), which is 10 times more massive than the SMC, has only 15 confirmed BeXRBs \citep{2013A&A...554A...1M}. A possible explanation is the connection between the density of BeXRBs and the recent star formation of the galaxy. \citet{2010ApJ...716L.140A} found a correlation between the SMC BeXRB population and the recent star formation $\sim(25-60)~\rm{Myr}$ ago. For the LMC a similar correlation has been reported but for stellar populations with ages $\sim(15-50)~\rm{Myr}$ \citep{2011AAS...21822829A}. The most recent star-formation episodes in the LMC occurred 12 and 100 Myr ago \citep{2009AJ....138.1243H}. The X-ray coverage of the LMC is not nearly as complete as that of 
the SMC and so early interpretations on the connection between star-formation rate and its BeXRB population would be dubious. The \xmm~ survey of the LMC (PI: F. Haberl) will achieve a homogeneous coverage of the inner $\sim4.5\degr\times4.5\degr$ region of the LMC and will provide us with a more complete BeXRB sample of this region.

In this work we report on a newly discovered BeXRB in the LMC, \swiftp. The source was discovered on 2011 Nov. 6 during a Swift observation of the LMC nova 2009B \citep{2011ATel.3747....1S}. It was then detected in a subsequent Swift follow-up observation that yielded an improved position and X-ray spectrum \citep{2011ATel.3753....1S}. Based on its spectral properties and the correlation with an early-type optical counterpart the source was characterised as a BeXRB candidate. Variability of the optical counterpart has been reported \citep{2011ATel.3751....1C}, which is consistent with such a classification. In addition to the existing Swift observations we present the spectral and timing analysis of an \xmm~ target-of-opportunity (ToO) observation as well as the results of optical spectroscopy performed on its companion.     

\section{Observations and data reduction}
 \label{sec-observations}

\subsection{X-ray observations}   
\label{sec-xobs}
For the current analysis we used six Swift, one \xmm~ and one ROSAT observations of the field surrounding the source. The detailed observation log is recorded in Table~\ref{tab:xray-obs}.

The source was detected in four more Swift observations of the field in addition to those reported above. For the spectral extraction we used the {\tt HEASoft}\footnote{http://heasarc.nasa.gov/lheasoft/} task {\tt xselect} and an extraction region with radius 35\arcsec~ for the source and 300\arcsec~ for the background. 

Following the detection of \swiftp~ in the Swift observation on 2012 Aug. 28, we obtained an \xmm~ ToO observation (PI: P. Maggi). The observation was performed on 2012 Sep. 03 and allowed us to detect the source during outburst. The source was located on-axis, on CCD4 of the EPIC-pn \citep{2001A&A...365L..18S} and on CCD1 of the EPIC-MOS \citep{2001A&A...365L..27T} detectors. \xmm~ SAS 12.0.1\footnote{Science Analysis Software (SAS), http://xmm.esac.esa.int/sas/} was used for data processing. The observation was affected by high background for the first 10 ks, which then stayed at a moderate level for the rest of the exposure. We used a background threshold of 16 and 8 counts ks$^{-1}$ arcmin$^{-2}$ for the EPIC-pn and EPIC-MOS detectors respectively which resulted in net exposure times of 22.3/26.0/26.0 ks for EPIC-pn/MOS1/MOS2 after flare removal. The source event extraction was performed using a circle around the source while the background events were extracted from a point-source free area on the 
same CCD 
but with different pixel columns for EPIC-pn. We optimised the size of the source extraction area using the SAS task {\tt eregionanalyse}. The final values used for the extraction are given in Table \ref{tab:xray-obs}, which summarises the available X-ray observations. For the EPIC-pn spectra and light curves, we selected single-pixel and double-pixel events ({\tt PATTERN$\le$4}) while in the case of EPIC-MOS we used single to quadruple events ({\tt PATTERN$\le$12}). The quality flag {\tt FLAG = 0} was used throughout. The SAS task {\tt especget} was used to create the spectra for the spectral analysis. The spectra were binned to achieve a minimum signal-to-noise ratio of five for each bin. This allows the use of the $\chi^2$ statistics in the spectral analysis, which was performed with {\tt XSPEC} \citep{1996ASPC..101...17A} version 12.7.0.

A ROSAT archival pointed observation covering the position of \swiftp~ is also available, which we used to establish an upper limit for the source luminosity.

\subsection{Optical Spectroscopy}
\label{sec-opt_spec}

The optical data were taken with the ESO Faint Object Spectrograph and Camera (EFOSC2) mounted at the Nasmyth B focus of the 3.6 m New Technology Telescope (NTT), La Silla, Chile on the nights of 2011 Dec. 8, 9 and 10. A slit width of 1.5\arcsec~ was used, along with Grisms 14 and 20 for blue and red end spectroscopy, respectively. For details on the individual Grisms see Table \ref{Tab:Grisms}. Filter OG530 was used in conjunction with Grism 20 to block second order effects. The resulting spectra were recorded on a  Loral/Lesser, thinned, AR-coated, UV flooded, MPP CCD with 2048$\times$2048 pixels. Wavelength calibration was achieved using comparison spectra of helium and argon lamps taken throughout the observing run with the same instrument configuration.
The data were reduced using the standard packages available in the Image Reduction and Analysis Facility ({\tt IRAF})\footnote{Image Reduction and Analysis Facility (IRAF), http://iraf.noao.edu/}. The spectra were normalised to remove the continuum and a redshift correction was applied corresponding to the recession velocity of the LMC (-280~km~s$^{-1}$, \citealt{1987A&A...171...33R}).

\begin{table}
\begin{center}
\caption{ESO Faint Object Spectrograph grism information.}\label{Tab:Grisms}
\scalebox{0.9}{\begin{tabular}{lcccc}
\hline\hline
Grism & Wavelength & Grating & Dispersion & Resulting \\
& range [\AA{}]& [lines~mm$^{-1}$] & [\AA{}~pixel$^{-1}$] & resolution [\AA{}]\\\noalign{\smallskip}
\hline\noalign{\smallskip}
14 & 3095--5085 & 600 & 1.00 & $\sim$10\\\noalign{\smallskip}
20 & 6047--7147 & 1070 & 0.55 & $ \sim$6\\ \noalign{\smallskip}
\hline
\end{tabular}}
\end{center}
\end{table}

\section{X-ray data analysis and results}
 \label{sec-datareduction_Xray}

\subsection{X-ray position}
X-ray images were created from all the EPIC cameras using the \xmm~ standard energy sub-bands \citep{2009A&A...493..339W}. Source detection was performed simultaneously on all the images using the SAS task {\tt edetect\_chain}. Boresight correction was performed on the images based on the position of three identified background X-ray sources in the same field. The positional correction based on these sources was found to be $\sim$ 1\arcsec. The source position was determined to R.A. = 05$^{\rm h}$30$^{\rm m}$42\fs17 and Dec. = --66\degr54\arcmin31\farcs0 (J2000.0) with a $1\sigma$ statistical uncertainty of 0.08\arcsec. The total 1 $\sigma$ positional uncertainty, however is determined by the remaining systematic uncertainty assumed to be 0.5\arcsec \citep[see section 4.3 of][]{2013arXiv1307.7594S}. 

\subsection{Spectral analysis}
\label{sec-spec_x}
All the EPIC spectra were fitted simultaneously for the same model parameters with an additional scaling factor to account for instrumental differences. For the EPIC-pn we fixed the scaling factor to be 1 while for both EPIC-MOS we obtained values of $C_{MOS}=1.08\pm0.05$, which is consistent with the expected value, as EPIC-MOS is known to provide $\sim5\%$ higher fluxes than EPIC-pn
\citep[see ][or the latest version of the \xmm~ calibration manual\footnote{http://xmm2.esac.esa.int/external/xmm\_sw\_cal/calib/cross\_cal/index.php}]{2006ESASP.604..937S}. The photoelectric absorption was modelled as a combination of Galactic foreground absorption and an additional column density accounting for both the interstellar medium of the LMC and the intrinsic absorption by the source. The Galactic photoelectric absorption was set to a column density of N$_{\rm H{\rm , GAL}}$ = 5.86\hcm{20} \citep{1990ARA&A..28..215D} with abundances according to \citet{2000ApJ...542..914W}. The additional column density $N_{\rm{H},\rm{LMC}}$ was left as a free parameter with abundances of 0.49  for elements heavier than helium \citep{2002A&A...396...53R}. All the uncertainties were calculated based on a 
 $\Delta\chi^2$ statistic of 2.706, equivalent to a 90\% confidence level for one parameter of interest. 

\begin{figure}
   \resizebox{\hsize}{!}{\includegraphics[angle=-90,clip=]{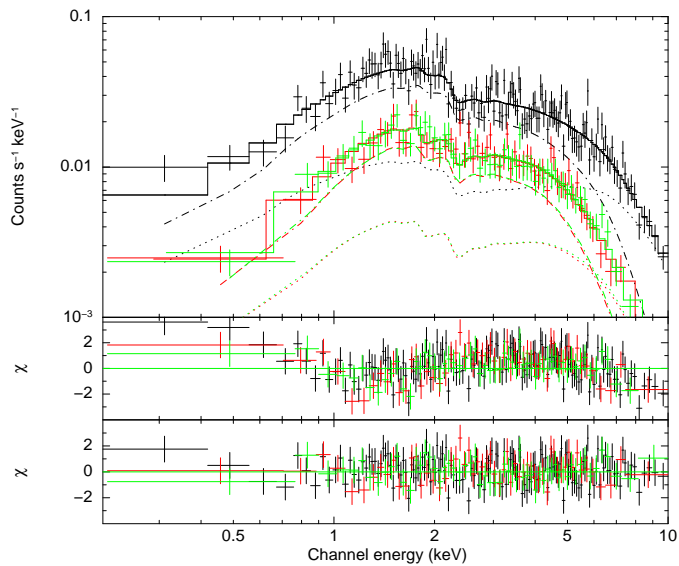}}
      \caption{EPIC spectra of \swiftp. The top panel shows the EPIC-pn (black), EPIC-MOS1 (red) and EPIC-MOS2 (green) spectra, together with the best-fit model (solid line) 
            of an absorbed power-law (dotted line) plus a black-body (dash-dotted line).
 	   The residuals for this model are plotted below (bottom panel) while for comparison the best-fit single power-law model residuals are also plotted (middle panel).}
   \label{spec}
 \end{figure}

The spectra were fitted by an absorbed power law, resulting in an acceptable fit with reduced $\chi^2$ value of $\chi^2_{\rm{red}}=1.4$. However, we found systematic deviations in the residuals. Therefore, we further tested the spectra for the existence of additional features typically shown by BeXRBs. The best fit was achieved by an absorbed power law plus a black body that contributes to the soft part of the spectrum. The two-component model significantly improves the fit over the simple power-law ($\chi^{2}_{\rm{red}}$=0.98 vs 1.4 for two additional parameters), with the F-test probability for the additional black-body component being $\sim8\times10^{-21}$. The black-body component accounts for 53.8\% of the detected flux and demonstrates model dependent uncertainties on the photon index of the power-law model. The black-body model {\tt bbodyrad} uses a normalisation proportional to the emission area (norm $K=R^{2}_{km}/D^{2}_{10}$, where $R_{km}$ denotes the emission radius in km and $D_{10}$ the source 
distance in units of 10 kpc). From this relation we derived a radius for the emission region of $\sim800$ m. This area should be indicative of the size of the accretion column from which the black-body radiation is believed to originate \citep{2004ApJ...614..881H}. The spectra together with the best-fit power-law and black-body model are presented in Fig.~\ref{spec}, where we also compare the residuals with those of the simple power-law model fit. 

\begin{figure}
   \resizebox{\hsize}{!}{\includegraphics{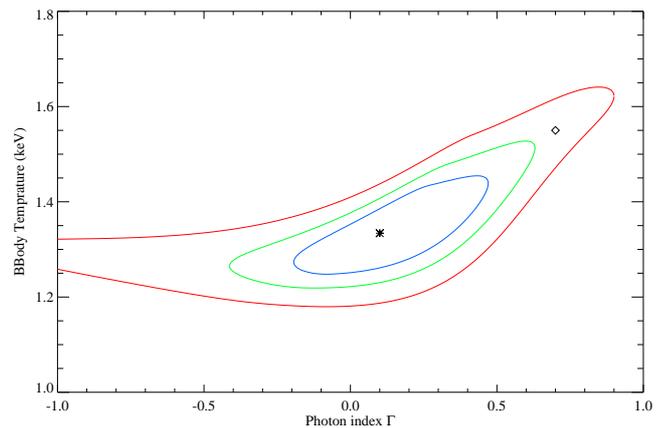}}
      \caption{Photon index $\Gamma$ vs temperature $kT~({\rm keV})$ confidence contours for the power-law + black-body model. Confidence levels of 1, 2 and 3 $\sigma$ are shown in blue green and red. The star represent the best-fit values while the diamond marks the best fit for a fixed $\Gamma$ of 0.7.}
   \label{contour}
 \end{figure}

It should be noted that although the addition of the black-body model improves the quality of the fit it also adds large uncertainties to the best-fit values. Thus different combinations of parameters can actually provide a similarly good fit (see Fig. \ref{contour}).  Softer power-laws (higher $\Gamma$, where $\Gamma$ is defined as the index in a power low spectrum $A(E)\propto E^{-\Gamma}$) can be compensated with higher absorption values, while for lower $\Gamma$ the best fit model gives lower absorption values and lower black-body temperature. Therefore, for any constant value of $\Gamma$ between 0 and 0.8 the reduced  $\Delta\chi^2$ for the fit varies only marginally around one. Also by looking at Fig. \ref{spec}, we can see that the black-body component contributes almost to all the emission below 6 keV. Because of the large uncertainties, we present another more consistent set of parameters for a fixed photon index at the value of 0.7, which is close to the one found with the single power-law fitting. 
The best-fit results are listed in Table~\ref{tab-spectra}. 

Replacing the black-body component by alternative models such as, a Comptonisation spectrum, a thermal bremsstrahlung or a MEKAL thermal plasma only gives a marginal improvement of the fit compared to the single power-law model. The only models that showed similar goodness of fit were a power-law with high-energy exponential cutoff ({\it cutoffpl} in {\it xspec}, with $\Gamma=-0.68$ and e-folding energy of 2.8 keV) and the broken power-law ({\it bknpower} in {\it xspec}, with $\Gamma_1=0.1$, $\Gamma_2=1.2$, break at 3.3 keV). However, we do not expect negative photon indices nor a break at such low energies for BeXRB systems. 

Individual Swift observations did not provide sufficient statistics to determine spectral parameters independently, but using a group of them to simultaneously fit a model provided better statistics. For the first four observations that were taken during the same outburst and were at similar flux level we performed the event extraction as described in Section \ref{sec-xobs} and performed a simultaneous spectral fitting with {\tt XSPEC}. Due to lower statistics compared to the \xmm~ spectra we only tested an absorbed power-law model. The best-fit values are also listed in Table \ref{tab-spectra}. For the two Swift observations taken in 2012 we used the best-fit parameters for the absorbed power-law model as obtained from the \xmm~ spectrum (near in time) and derived the fluxes of these Swift observations by adjusting the normalisation.

From the ROSAT observation we estimated an upper limit for the X-ray flux assuming the same spectral properties as in the \xmm~ detection and assuming that for a detection at least seven counts are needed for an on-axis observation.

\begin{table*}
\caption[]{Spectral fit results.}
\begin{center}
\begin{tabular}{lcccccccc}
\hline\hline\noalign{\smallskip}
\multicolumn{1}{l}{Model$^{(1)}$} &
\multicolumn{1}{c}{LMC \nh} &
\multicolumn{1}{c}{$\Gamma$} &
\multicolumn{1}{c}{kT} &
\multicolumn{1}{c}{R$^{(2)}$} &
\multicolumn{1}{c}{Flux$^{(3)}$} &
\multicolumn{1}{c}{L$_{\rm x}^{(4)}$} &
\multicolumn{1}{c}{L$_{\rm BB}^{(5)}$} &
\multicolumn{1}{c}{$\chi^2_{\rm{red}}/{\rm dof^{(6)}}$} \\
\multicolumn{1}{c}{} &
\multicolumn{1}{c}{[\oexpo{21}cm$^{-2}$]} &
\multicolumn{1}{c}{} &
\multicolumn{1}{c}{[keV]} &
\multicolumn{1}{c}{[m]} &
\multicolumn{1}{c}{[erg cm$^{-2}$ s$^{-1}$]} &
\multicolumn{1}{c}{[erg s$^{-1}$]} &
\multicolumn{1}{c}{[$\%$]} &
\multicolumn{1}{c}{} \\

\noalign{\smallskip}\hline\noalign{\smallskip}
\xmm~ & & & & & & & \\
PL               & $5.8^{+0.1}_{-0.09}$               & 0.78$\pm$0.06       & --                     & --                   & 2.07\expo{-12} & 5.5\expo{35} & --    &  1.40/272 \\
PL+BBrad         & $0^{+1.1}$   & $0.1^{+0.4}_{-0.3}$ & $1.33^{+0.14}_{-0.09}$ & $850^{+100}_{-130}$  & 1.79\expo{-12} & 5.3\expo{35} & 53.8  &  0.98/270 \\
PL(fixed)+BBrad  & $1.4\pm0.8$                        & 0.7 fixed           & $1.55^{+0.06}_{-0.07}$  & $680\pm60$           & 1.78\expo{-12} & 5.3\expo{35} & 60.0    & 1.00/271 \\
\hline
Swift$^{(7)}$ & & & & & & & \\
PL  & $1.0^{+4.4}_{-1.0}$& $1.14^{+0.09}_{-0.08}$ & -- & -- & -- & --  & --& 0.90/79 \\
\noalign{\smallskip}\hline
\end{tabular}
\end{center}
Notes:$^{(1)}$ For definition of spectral models see text. 
$^{(2)}$ Radius of the emitting area.
$^{(3)}$ Observed (0.2-10.0) keV flux derived from EPIC-pn.
$^{(4)}$ Source intrinsic X-ray luminosity in the (0.2-10.0) keV band (corrected for absorption)
for a distance to the LMC of 50 kpc \citep{2005MNRAS.357..304H}.
$^{(5)}$ Contribution of the black-body component to the total unabsorbed luminosity.
$^{(6)}$ Degrees of freedom (dof).
$^{(7)}$ Best-fit parameters for simultaneous fitting of the spectra from the first four Swift observations.  
\label{tab-spectra}
\end{table*}

\subsection{X-ray long-term variability}
\label{sec-Xvar}

In order to study the long-term variability of \swiftp~ we estimated the flux level for all available observations (see Table \ref{tab:xray-obs}). The source flux from the \xmm~ ToO observation was determined from the spectra as described in Section \ref{sec-spec_x}. We transformed the observed fluxes to unabsorbed luminosities based on a LMC distance of 50 kpc. The net count rates and unabsorbed luminosities are given in Table \ref{tab:xray-obs}.

The Swift observations indicate a variability of up to a factor of $\sim14$. The \xmm~ ToO observation yields a luminosity lower than any Swift observation giving a total variability of $\sim23$. The upper limit derived 
from the ROSAT observation is $\sim$1.6 times smaller than the \xmm~ detected flux, providing evidence for even larger variability. However, the photoelectric absorption will have large effect on the soft ROSAT energy band and may introduce large systematic uncertainties.

\subsection{Timing analysis}
\label{sec-time_an}

\begin{figure}
 \resizebox{\hsize}{!}{\includegraphics[angle=-90,clip=]{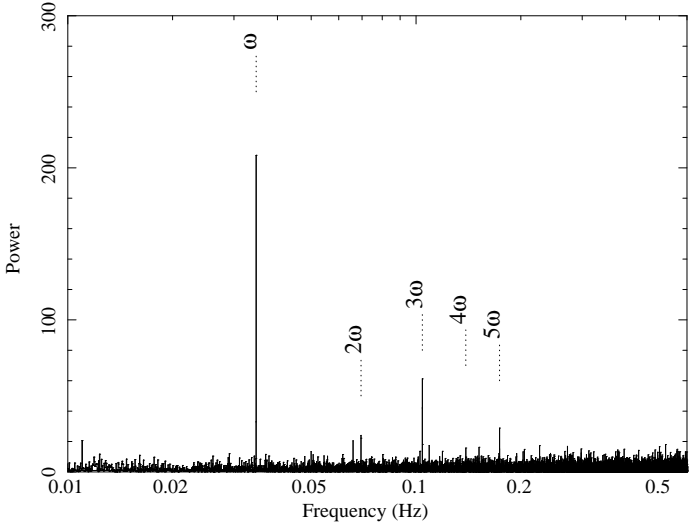}}
  \caption{Power density spectrum created from the merged EPIC-pn and EPIC-MOS data in the (0.2-10.0) keV energy band. 
           The time binning of the input light curve is 0.8156671 s. On the plot we can see the best-fit frequency of $\omega=0.034752$ Hz and its first, second and fourth harmonics.}
  \label{fig:psd}
\end{figure}

\begin{figure}
  \resizebox{\hsize}{!}{\includegraphics[angle=0,clip=]{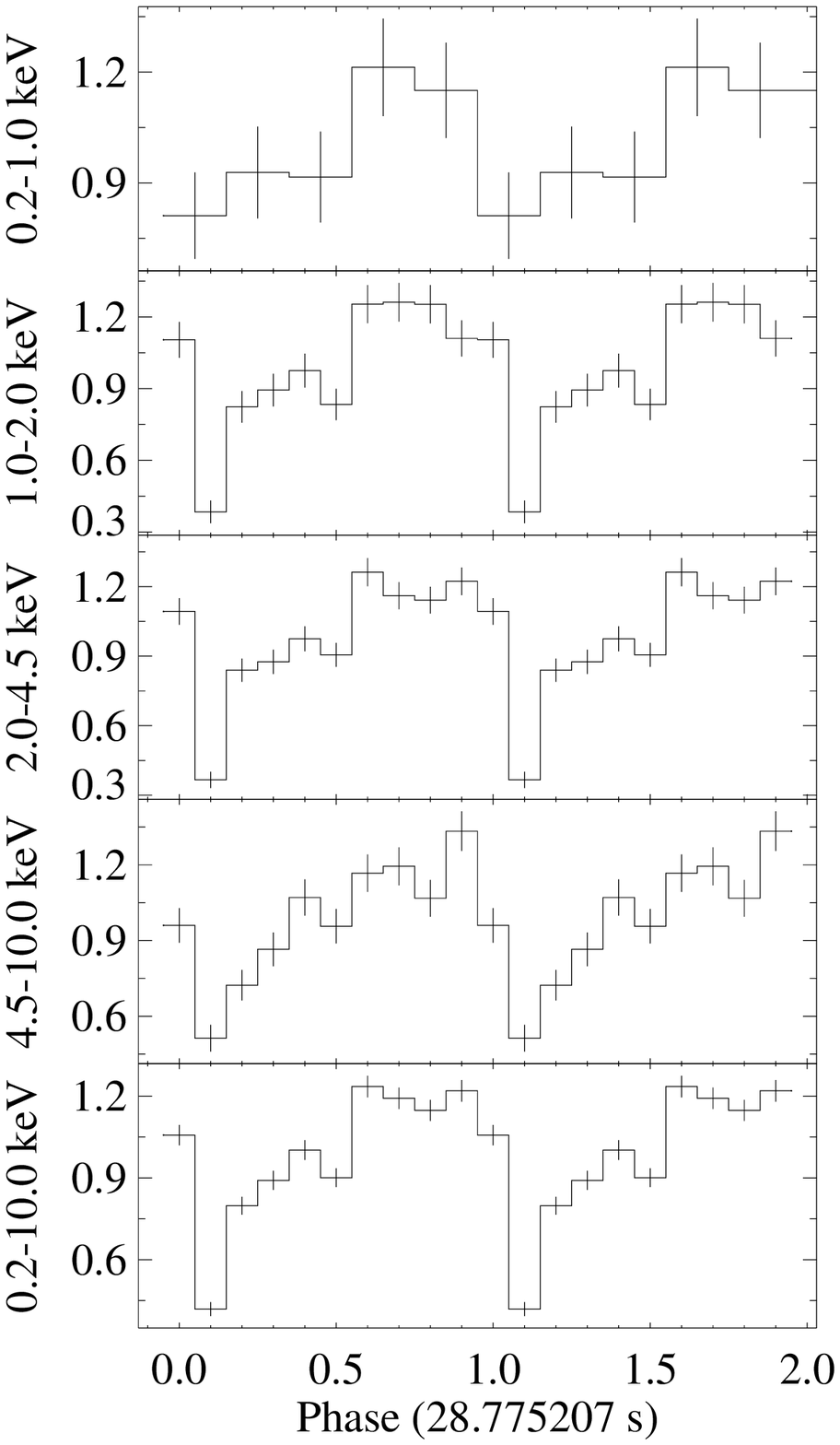}\includegraphics[angle=0,clip=]{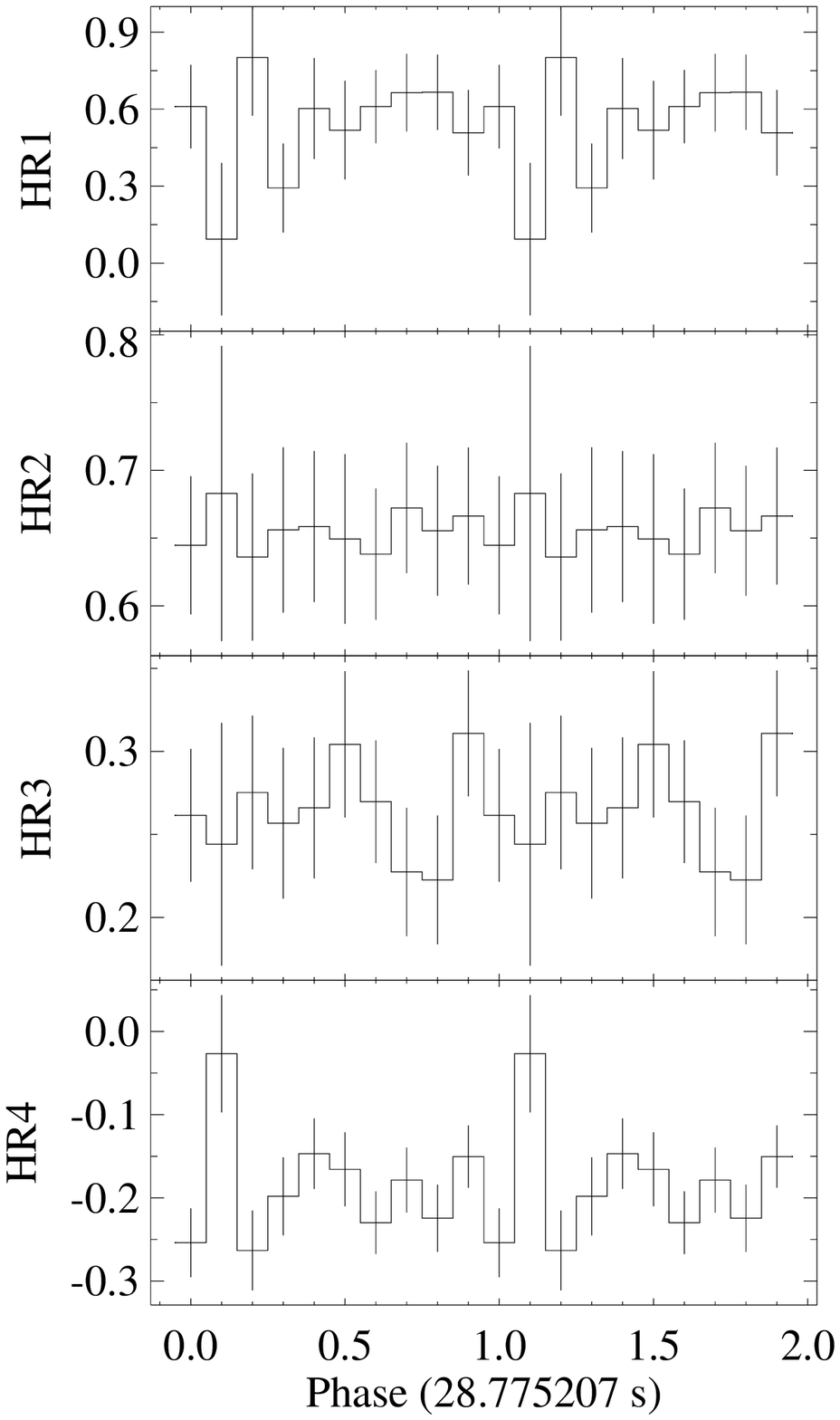}}
  \caption{Left: Pulse profiles obtained from the merged EPIC data in different energy bands.
Right: Hardness ratios as a function of pulse phase derived from the pulse profiles in two neighbouring standard energy bands. 
	  }
  \label{fig:pp}
\end{figure}

We corrected the \xmm~ EPIC event arrival times to the solar-system barycentre using the SAS task {\tt barycen}. The X-ray light curves were searched for periodicities using fast Fourier transform (FFT) and light curve folding techniques.
The power density spectra derived from time series in various energy bands from all three EPIC instruments show a periodic signal at $\omega\sim0.0347521$ Hz. 
To increase the signal-to-noise ratio, we then created time series from the merged event list of EPIC-pn and EPIC-MOS (delimited to common good time intervals).
Fig.~\ref{fig:psd} shows the inferred power density spectrum from the (0.2-10) keV energy band with a strong peak at a frequency of 0.0347520(5)~Hz.
Following \citet{2008A&A...489..327H} we used a Bayesian periodic signal detection method \citep{1996ApJ...473.1059G} to determine the pulse period 
with 1$\sigma$ uncertainty to $(28.77521\pm 0.00010)$ s. The period folded pulse profiles in the EPIC standard energy bands together with the hardness ratios derived from the pulse profiles in two adjacent energy bands ($\rm{HR}_i=(\rm{R}_{\rm{i+1}}-\rm{R}_{\rm{i}})/(\rm{R}_{\rm{i+1}}+\rm{R}_{\rm{i}})$ with R$_{\rm i}$ denoting the background-subtracted count rate in energy band i) are plotted in Fig. \ref{fig:pp}. To achieve better statistics, the first two standard energy bands were combined in the top panel, the bottom panel shows all five energy bands combined. All the profiles are background-subtracted and normalised to the average count rate (0.0215, 0.0813, 0.1403, 0.0950, and 0.3388 \ct, from top to bottom).

\section{Analysis and results of optical data}
\label{datareduction_optical}

\begin{figure}
\resizebox{\hsize}{!}{\includegraphics[clip=]{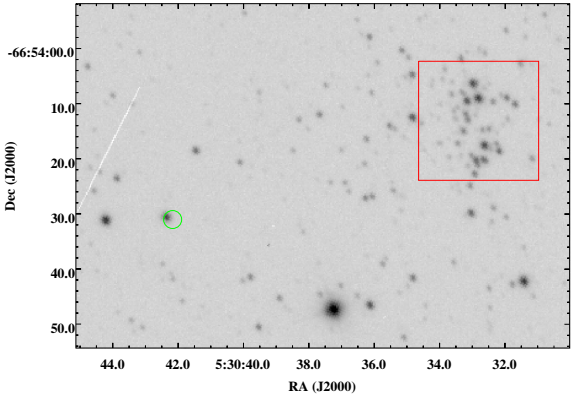}}
  \caption{Finding chart of \swiftp. The image was taken with ESO NTT (V\#641 filter) during the spectroscopic observation of the source. The green error circle with radius of 1.6\arcsec~ translates to the $3\sigma$ \xmm~  positional uncertainty. On the right part of the image at a distance of 0.95\arcmin~ from the \xmm~ position of \swiftp~ the nearby star cluster KMHK 987 is seen (dashed red box).}
  \label{fchart}
\end{figure} 

\subsection{Optical counterpart}
The most precise X-ray position of \swiftp~ was derived from \xmm~ data (Sect. \ref{sec-xobs}). Within the 3$\sigma$ error circle we found only one candidate counterpart at a distance of $\sim$0.8\arcsec. The object was identified in 2MASS \citep{2006AJ....131.1163S,2003yCat.2246....0C}, IRSF \citep{2007PASJ...59..615K} and MCPS \citep{2004AJ....128.1606Z} catalogs and the WISE All-sky data release. It is listed as J05304215-6654303 in the 2MASS catalogue. The various magnitudes are summarised in Table \ref{tab:opt-mag}. 

A finding chart of \swiftp~ is shown in Fig. \ref{fchart}. The image was obtained with ESO NTT during the spectroscopic observation of the source and the image coordinates were corrected based on the 2MASS catalogue with a positional uncertainty $\sim0.2$\arcsec. 

The spectral energy distribution (SED) of Be stars consists of two components \citep{1994A&A...290..609D}, the emission from the surface of the star (stellar atmosphere of T$\sim~(18\,000-38\,000)~{\rm K}$) and an infrared excess from the equatorial disc. Due to the evolution of the disc with time the luminosity of the system may vary. Therefore, simultaneous measurements of individual colours are needed for a meaningful SED. From the available data it is not possible to construct an SED and compare it with available models. However, we see evidences for variability of the optical counterpart, which is typical for BeXRB systems. 

The optical counterpart of \swiftp~ was also found in the {\it Spitzer SAGE catalogs of the Magellanic Clouds} \citep{2006AJ....132.2268M}. Using the J colours from various surveys and the IRAC 3.6, 4.5, 5.8 $\mu$m colours (see table \ref{tab:opt-mag}) and based on the analysis of \citet{2009AJ....138.1003B,2010AJ....140..416B} we determined the position of the source on the  $J_{\rm IRSF}$ vs. $J_{\rm IRSF}$ - [3.6], $J_{\rm IRSF}$ vs. $J_{\rm IRSF}$ - [5.8], $J_{\rm IRSF}$ vs. $J_{\rm IRSF}$ - [8.0] diagrams. Most of the B stars have a $J$ - [3.6] colour $\sim[-0.2:0.2]$ mag. While, depending on the four available observations in the J filter, the optical counterpart of \swiftp~ has a larger colour index between  0.7 and 1.26 mag. Based on that we clearly detect infrared excess of the counterpart of \swiftp. This excess can be attributed to the free-free emission from the disc around the star.

\begin{table*}
\caption{Optical and IR photometry of \swiftp.}
\begin{center}
\begin{tabular}{ccccccccc}
\hline\hline\noalign{\smallskip}
   & Z2004$^{(a)}$  & 2MASS$^{(b)}$   & 2MASS 6X$^{(c)}$   & 2MASS 6X$^{(c)}$ & IRSF$^{(d)}$&  WISE$^{(e)}$ & SAGE$^{(f)}$ \\
   & 2004     & 2000 Feb 28     & 2000 Dec 8   & 2001 Feb 4 & 2003 Jan 28 &- & -\\
  MJD\, & -     & 51602.0488      & 51886.3355  & 51944.1069  & 52667.896  & - & -  \\
\noalign{\smallskip}\hline\noalign{\smallskip}
  U & 14.312$\pm$0.022 & $-$ & $-$ & $-$ & $-$ & $-$ & $-$ \\
  B & 15.007$\pm$0.033 & $-$ & $-$ & $-$ & $-$ & $-$ & $-$  \\
  V & 15.321$\pm$0.018 & $-$ & $-$ & $-$ & $-$ & $-$ & $-$  \\
  I & 15.586$\pm$0.023 & $-$ & $-$ & $-$ & $-$ & $-$ & $-$ \\
  J & $-$ & 15.26$\pm$0.05 & 15.83$\pm$0.04 & 15.66$\pm$0.04 & 15.30$\pm$0.02  & $-$ & $-$ \\
  H & $-$ & 14.98$\pm$0.09 & 15.82$\pm$0.09 & 15.86$\pm$0.07 & 15.24$\pm$0.02  & $-$ & $-$\\
  K & $-$ & 14.91$\pm$0.13 & 15.86$\pm$0.16 & 15.85$\pm$0.13 & 15.17$\pm$0.05  & $-$ & $-$\\
  W1 & $-$ & $-$ & $-$ & $-$ & $-$ & 14.245$\pm$0.027 & $-$ \\
  W2 & $-$ & $-$ & $-$ & $-$ & $-$ & 14.125$\pm$0.027 & $-$ \\
  W3 & $-$ & $-$ & $-$ & $-$ & $-$ & <13.898$^{(g)}$  & $-$ \\
  W4 & $-$ & $-$ & $-$ & $-$ & $-$ & <10.086$^{(g)}$  & $-$\\
 $[3.6]$ & $-$ & $-$ & $-$ & $-$ & $-$ &  $-$ & 14.566$\pm$0.039 \\
 $[4.5]$ & $-$ & $-$ & $-$ & $-$ & $-$ &  $-$ & 14.343$\pm$0.035 \\
 $[5.8]$ & $-$ & $-$ & $-$ & $-$ & $-$ &  $-$ & 14.247$\pm$0.048 \\
 
\noalign{\smallskip}\hline\noalign{\smallskip}
\multicolumn{5}{l}{\hbox to 0pt{\parbox{180mm}{\footnotesize
\smallskip
$^{(a)}$\citet{2004AJ....128.1606Z}. 
$^{(b)}$\citet{2003yCat.2246....0C}. 
$^{(c)}$\citet{2012yCat.2281....0C}.
$^{(d)}$\citet{2007PASJ...59..615K}. 
$^{(e)}$\citet{2012yCat.2311....0C}.
$^{(f)}$\citet{2006AJ....132.2268M}.
$^{(g)}$ upper limits.
}}}
\end{tabular}
\end{center}
\label{tab:opt-mag}
\end{table*}

\subsection{Spectral Classification}

\begin{figure}
\centering
\resizebox{\hsize}{!}{\includegraphics[clip=]{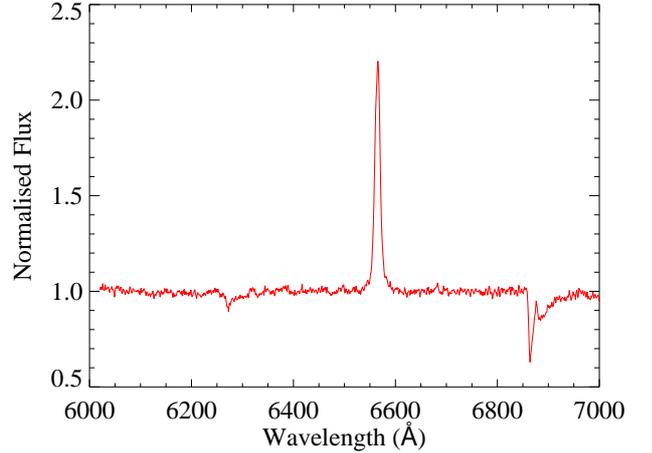}}
\caption{Spectrum of \swiftp~ in the wavelength range $\lambda$6000--7000\AA{} as measured with the NTT on 2011-12-10. The spectrum has been normalised to remove the continuum and redshifted by $-280$~km~s$^{-1}$.}\label{fig:red}
\end{figure}

From the optical spectroscopy we clearly see H$\alpha$ emission from the proposed counterpart of \swiftp~ . Fig. \ref{fig:red} shows the red end of the spectrum taken on 2011 Dec. 10. The H$\alpha$ equivalent width, considered as an indicator for circumstellar disc size \citep{1997MNRAS.288..988S}, is $-(16.8\pm0.9)$~\AA{}.

OB stars in our own Galaxy are classified using the ratio of certain metal and helium lines \citep{1990PASP..102..379W} based on the Morgan-Keenan (MK; \citealt{1943QB881.M6.......}) system. However, in lower metallicity environments such as the Magellanic Clouds this is not possible, due to the absence or weakness of the metal lines. Therefore, the optical spectrum of \swiftp~ was classified using the method developed by \citet{1997A&A...317..871L} for B-type stars in the SMC. The luminosity classification method from \citet{1990PASP..102..379W} was assumed in this work.

\begin{figure*}
\centering
\includegraphics[height=180mm,angle=90]{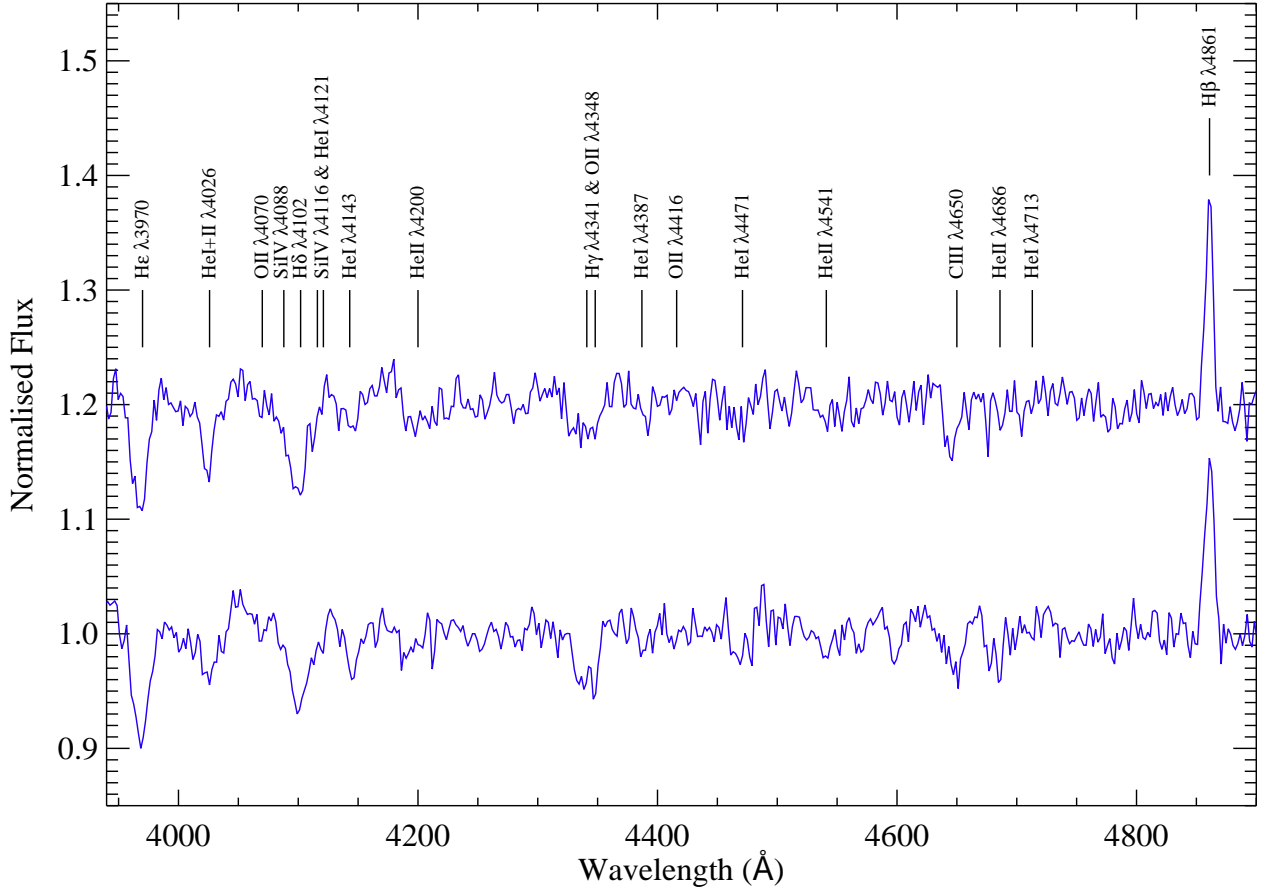}
\caption{Spectra of \swiftp~ in the wavelength range $\lambda$3900--5000\AA{} taken on concurrent nights with the NTT. The spectra have been normalized to remove the continuum and redshift corrected by -280~km~s$^{-1}$. Atomic transitions relevant to spectral classification have been marked. The lower one was taken on the first night while a different normalization has been used for better representation and comparison.}\label{fig:blue}
\end{figure*}

Figure \ref{fig:blue} shows two unsmoothed optical spectra of \swiftp~ taken on 2011 Dec. 8 and 9, dominated by the hydrogen Balmer series. The equivalent widths of the H$\beta$ lines are $(-1.3\pm0.2$)~\AA{} and $(-1.4\pm0.2$)~\AA{}, respectively. The Si\textsc{iv} lines at $\lambda4088,\lambda4116$~\AA{} are both clearly present in both spectra, despite the rotational broadening of the H$\delta$ line, suggesting a classification earlier than B1.5. The He\textsc{ii} line at $\lambda4686$~\AA{} is just visible above the noise level of the data, implying a classification of B1. There also appears to be marginal evidence for a line at $\lambda4541$~\AA{} (particularly in the lower spectrum) indicating a spectral class of B0.

The luminosity class of the system was determined using the ratios of \ion{Si}{iv}$\lambda4088$/\ion{He}{i+ii}$\lambda4026$,  \ion{Si}{iv}$\lambda4088$/\ion{He}{i}$\lambda4143$, \ion{Si}{iv}$\lambda4116$/\ion{He}{i}$\lambda4121$ and \ion{He}{ii}$\lambda4686$/\ion{He}{i}$\lambda4713$. The first three increase with increasing luminosity (decreasing luminosity class) and the latter increases with decreasing luminosity (increasing luminosity class). All but the Si\textsc{iv}$\lambda4116$/He\textsc{i}$\lambda4121$ ratio point to a luminosity class V (the other suggesting a luminosity class III). This is supported by the weak O\textsc{ii} spectrum, characteristic of dwarf stars. The \emph{V}-band magnitude of the companion star is reported to be $(15.32\pm0.02)$ mag by \citet{2004AJ....128.1606Z}. Along with a distance modulus of $(18.52\pm0.07)$ mag \citep{2011ApJ...729L...9B} and an $A_V$ of $(0.26\pm0.02)$ mag calculated using the Galactic column density towards the source of $(5.8\pm0.
3)\times 10^{20}
$~cm$^{-2}$ \citep{1990ARA&A..28..215D} and the results of \citet{2009MNRAS.400.2050G}, this leads to an absolute magnitude $M_V=-3.46\pm0.08$ mag, consistent with a B0V star. As such we classify the optical counterpart of \swiftp~ as a B0-1.5V star.

\subsection{MACHO light curve}

\begin{figure} 
\resizebox{\hsize}{!}{\includegraphics[clip=]{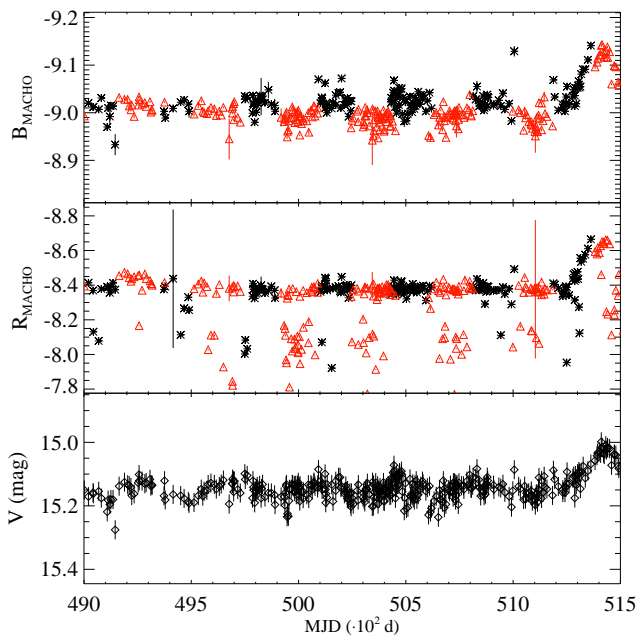}}
\caption{MACHO light curves for a time period of 2541 days: the ``blue'' $B_{\rm MACHO}$ lightcurve is plotted on the top panel, the ``red'' $R_{\rm MACHO}$ in the middle. In both of them red triangles and black stars indicate instrumental magnitudes taken on different pier setup. The corrected Johnson V magnitude is plotted on the bottom panel.}\label{fig:macho}

\end{figure}

We identified the optical counterpart in the on-line catalogue\footnote{http://wwwmacho.anu.edu.au/} (photometry id-61:8197:41:64925) of the MACHO (MAssive Compact Halo Objects) project. The MACHO observations were made using the 1.27-m telescope at Mount Stromlo Observatory, Australia. A dichroic beamsplitter and filters provide simultaneous CCD photometry in two passbands, the $R_{\rm MACHO}$ ’red’ band ($6300-7600$~\AA{}) and the $B_{\rm MACHO}$ ’blue’ band ($4500-6300$~\AA{}). The standard MACHO instrumental magnitudes are extracted with {\tt SoDoPHOT} and are based on point-spread function fitting and differential photometry relative to bright neighbouring stars. SoDoPHOT stands for Son of DoPhot, a revised package based on DoPhot\footnote{\texttt{software:} http://www.astro.puc.cl:8080/astropuc/software-hecho-en-daa} algorithms but optimized to MACHO image data. The {\tt SoDoPHOT} instrumental magnitude is given as 2.5 times the common logarithm of the integrated number of electrons in the fit to the 
analytical PSF, divided by 100. 
MACHO light curves may show systematic variations in brightness due to the different air-mass during individual observations. Additionally, artificial ``seasonal rolls'' ($\sim1~ {\rm yr}$ periods) may be a result of the target being observed at progressively higher air-masses as the season progresses. Finally instrumental light curves are affected by observations made from either side of the telescope pier (a German equatorial mount). Thus stars will alternatively be located on the CCDs rotated by 180 degrees from each other in the focal plane.                     
                                                                                                                                                                                                                                                                                                                                                                                                                                                                                                                                                                                            
In order to use the light curves we need to correct them for the above effects, e.g. the air-mass and the responses of the different CCDs. We followed the analysis of \citet{1999PASP..111.1539A}. In their work they derived corrections for all the instrumental magnitudes based on a statistical analysis of the properties of the detected stars in the MACHO fields of the LMC. But the relations given in their work cannot be applied in a straightforward way to a single star. In particular the coefficients for the air-mass-colour corrections must be derived using separate knowledge of the shape of a light curve \citep[e.g. equations 7 and 8 in][]{1999PASP..111.1539A}. In order to correct the instrumental light curve, we searched all the parameter space for these coefficients and derived the best solution for a continuous light curve. Finally we transformed the corrected magnitudes to the Johnson V and Cousins R bands. The complete analysis added a systematic uncertainty of $\sim0.02~{\rm mag}$  to the light curve. 

In Fig. \ref{fig:macho} we show the light curves for $R_{\rm MACHO}$, $B_{\rm MACHO}$ and the corrected Johnson V magnitude. The $B_{\rm MACHO}$ light curve shows evidence for periodicity and both instrumental magnitudes show an outburst at the end of the light curve. The magnitude of the outburst was 0.15 and 0.2 mag in the $B_{\rm MACHO}$ and the $R_{\rm MACHO}$ filter, respectively. 
The reddening of the source during the outburst is typical of Be star in the phase of growing  the circumstellar disc \citep{2006A&A...456.1027D}.
The $R_{\rm MACHO}$ light curve shows several dips up to 0.6 mag, that are more often seen in one of the instrument set-up. These dips are mostly seen during the winter and spring period and might be caused by weather conditions, which affect the $R_{\rm MACHO}$ filter more. Another characteristic of the dips is that they seem to appear more often in observations at low airmass values $<1.25$. Consequently we believe that these dips are only instrumental artifacts and decided to exclude them from our analysis.     

We computed the Lomb-Scargle (LS) periodogram \citep{1982ApJ...263..835S} for the light curves of the instrumental magnitudes, excluding the outburst (MJD>51150). For $B_{\rm MACHO}$ we found a period of 373 d which could be artificial as the instrument setup changes approximately every 6 months. For the $R_{\rm MACHO}$ we find a period of $\sim~316~{\rm d}$. After correcting the instrumental magnitudes, we get a period of $\sim390~{\rm d}$ for the V-band light curve with a LS power less than 15. The period is still close to one year. In order to test if all the corrections were done accurately we used the same technique to correct the light curves of other nearby stars. None of the other stars showed a period close to one year or showed a similar light curve.  We corrected the light curves of the source by normalising the two data sets obtained by different instrument/pier orientation by their average values, this caused the LS power to fall dramatically. If the period was intrinsic to the sources, we would 
expect the LS power to rise, not to fall. We would also expect a genuine period to be much more consistent across the different bands. Therefore we conclude that from the light curve we have no clear evidence for the optical period of the optical counterpart of \swiftp, in contrast of what the usage of the uncorrected  instrumental data might yield.

\section{Discussion}
 \label{discussion}

The X-ray fluxes (see Table \ref{tab:xray-obs}) suggest that \swiftp~ was detected during an outburst in Nov. 2011, while the \xmm~ ToO observation was performed at a lower luminosity level. The flux upper limit derived from the ROSAT observation is close to the source luminosity during the \xmm~ ToO observation, suggesting that the non-detection might be explained by high absorption. However, due to the nature of the source, we cannot rule out the possibility that the source was inactive during the ROSAT observation. There are examples of BeXRBs being inactive for a long period before brightening up again \citep[e.g. SAXJ0103.2-7209,][]{2008A&A...491..841E}.

Knowing that the BeXRB population correlates with star formation regions with ages $\sim15-50~\rm{Myr}$ \citep{2011AAS...21822829A} it is likely that the system originated from a nearby cluster.
Swift\,J053041.9-665426 is located near the star cluster KMHK 987 at a projected angular distance of 0.95\arcmin~(see Fig. \ref{fchart}). KMHK 987 has a radius of 0.48\arcmin~ and an age of $\sim$31 Myr \citep{2010A&A...517A..50G}. For the case of HMXRBs in the SMC a convincing link between the binary systems and near star clusters has been reported, that leads to an average space velocity of these systems arising from a supernova kick \citep{ 2005MNRAS.358.1379C}. Assuming a value of 5 Myr for the most likely maximum lifetime of the Be star after the NS has been formed \citep{1977ApJ...214L..19S} we can estimate a lower limit on the runaway velocity imposed to the system by the  supernova kick. 
In our case this leads to a velocity of 2.6 \kms. Knowing that the systems true motion has a random direction to our line of sight the true velocity could be much larger. The true velocity given as function of the angle $\theta$ between the line of sight and the moving direction of the source is {$u=2.6/\sin{\theta}$} \kms. On a statistical basis this value is smaller than the average velocity reported for BeXRB systems in our galaxy \citep[e.g. $19\pm8$ \kms,][]{2000A&A...364..563V}. A simple explanation could be that the value we used for the time since the creation of the binary is arbitrary and could be much smaller, or that the system could originate from a more distant cluster. Another close cluster NGC 2002, which is even younger with age $\sim11$ Myr and is found at a distance of 2.55\arcmin~ ($\sim$37 pc). By doing the same calculations on this cluster we derive a lower limit for the runaway velocity of 14 \kms. Based on the above estimates, \swiftp~ could originate from either of those two clusters.

The power-law photon index derived from the EPIC spectra of 0.1--0.8 (depending on the model, and the existence of a black-body component) is on the hard side of the distribution of photon indices of BeXRBs, which has a maximum at $\sim0.9-1.0$ \citep{2008A&A...489..327H}. A spectral analysis of the available Swift data gives a higher index of $1.7\pm0.5$ \citep{2011ATel.3747....1S} and $1.15\pm0.17$ \citep{2011ATel.3753....1S}, but in both cases the statistics are much lower than that in the EPIC observation. 

The X-ray spectrum is well described with a combination of a power-law and a thermal black-body component, accounting for $\sim53\%$ of the flux in the (0.3-10.0) keV band. This soft component has been recently reported in many other BeXRB pulsars \citep[e.g.][]{2006A&A...455..283L,2012MNRAS.425..595R}
. From the parameters of the model, we estimate a black-body radius of $\sim800$ m for the emission region (accretion column). This value is in agreement with other persistent HMXRBs found in our Galaxy \citep{2013arXiv1301.5120L}. In fact we can compare the emission region size with the one estimated by the analysis of \citet{2004ApJ...614..881H}. Assuming  standard values for the neutron star mass, radius and magnetic field of $M_{\rm NS}=1.4~M_{\sun}$, $R_{\rm NS}=10^6$ cm and $B_{\rm NS}=10^{12}$ G, and by using the black-body luminosity derived by the fit we get an estimate of $\sim320$ m for the black-body emitting radius. If this description is correct we might expect some variability on the thermal component with the phase of the spin period, that could result in a variable hardness ratio. In the pulse profiles of \swiftp~ we see a 
significant change in the hardness ratio of the harder bands ($HR4$). This behaviour has been well studied in Swift\,J045106.8-694803, another BeXRB in the LMC \citep{2013subm}. Unfortunately, for \swiftp~ this occurs only for a small time interval and cannot provide good statistics for further investigation of the spectrum.

We note that one should be careful in the interpretation of the two-component model with power-law and black body. Insufficient statistics of the spectral data can lead to strong dependences of the photon index, black-body temperature and absorption column density. For example \citet{2009A&A...505..947L} derive a large negative $\Gamma$ for RX J1037.5-5647, a known BeXRB pulsar in the Milky Way, when using this model.

Some low strength peaks that are found in the optical periodograms for near one year period can be attributed both to the instrumental setup and weather observing conditions and thus prevent us from deriving a clear periodic behaviour, which could indicate an orbital period. 

\section{Conclusion}
\label{conclusion}
Analysis of the \xmm~ ToO observation of \swiftp~ revealed X-ray pulsations at $\sim28.78$~s. The X-ray spectrum is best described by a double component model, composed of an absorbed power-law and black-body. We have spectroscopically classified the optical counterpart as a B0-1.5Ve star. The strong IR signal, combined with the clear evidence for significant $H_\alpha$ emission, points convincingly to the existence of a circumstellar disc as the source of accretion material in the system. The analysis of the optical light curve showed a $\sim150$ d burst, but there is no clear optical period found in the two colour light curves. Our results confirm \swiftp~ as a BeXRB with a neutron star primary, making it the 16th known BeXRB pulsar in the LMC.

\begin{acknowledgements}
The \xmm\ project is supported by the Bundesministerium f\"ur Wirtschaft und
Technologie\,/\,Deutsches Zentrum f\"ur Luft- und Raumfahrt (BMWi/DLR, FKZ 50 OX
0001) and the Max-Planck Society. 
We thank the SWIFT team for accepting and carefully scheduling the target of opportunity observations, and we acknowledge the use of public data from the SWIFT data archive. 
This paper utilizes public domain data obtained by the MACHO Project, jointly funded by the US Department of Energy through the University of California, Lawrence Livermore National Laboratory under contract No. W-7405-Eng-48, by the National Science Foundation through the Center for Particle Astrophysics of the University of California under cooperative agreement AST-8809616, and by the Mount Stromlo and Siding Spring Observatory, part of the Australian National University.
G.V., P.M. and R.S acknowledge support from the BMWi/DLR grant FKZ 50 OR 1208, 1201 and 0907 respectively.
\end{acknowledgements}

 \bibliography{general}

\end{document}